\documentclass[doublespacing]{elsart}
\usepackage{icarus}

\usepackage{natbib}

\usepackage{longtable}
\usepackage{lscape}
\usepackage{setspace}
\usepackage{lineno}
  
\bibpunct{(}{)}{;}{a}{,}{,}

\usepackage{graphicx}
\usepackage{amssymb}
\usepackage{wasysym}

\newcommand{\BibTeX}{ \textrm{B\kern-.05em\textsc{i\kern-.025em b}\kern-.08em
    T\kern-.1667em\lower.7ex\hbox{E}\kern-.125emX} }

\begin{document}

\begin{frontmatter}



\title{Colors of Dynamically Associated Asteroid Pairs}


\author[dtm]{Nicholas A. Moskovitz}

\address[dtm]{Carnegie Institution of Washington, Department of Terrestrial Magnetism, 5241 Broad Branch Road, Washington, DC 20008 (U.S.A)}

%% This copyright statement isn't required at any stage by the Icarus
%% Editorial Office or Elsevier.  However, until you sign over the
%% copyright to Elsevier prior to publication (or negotiate with them
%% about copyright), you own the copyright to anything you create.
%% Just to keep things unambiguous, always include a copyright statement
%% or explicitly dedicate your work to the public domain.
\begin{center}
\scriptsize
Copyright \copyright\ 2012 Nicholas A. Moskovitz
\end{center}

%% ----- ELSEVIER STUFF -----
%% The commands below up to the \end{frontmatter} are commented out
%% so that we can do some Icarus-required formatting on the second and
%% third pages that is not required later on by Elsevier.  So when
%% your paper gets accepted, and you are ready to start dealing with
%% Elsevier, copy your abstract and keywords up here, uncomment these
%% lines, and comment out the ICARUS STUFF below.
%% 
%% Alternately, you might just want to move these abstract, keyword,
%% and end frontmatter commands down, and comment out the ICARUS STUFF
%% commands.  It doesn't matter.

% \begin{abstract}
% % Text of abstract
% 
% \end{abstract}
% 
% \begin{keyword}
% % keywords here, in the form: keyword \sep keyword
% 
% 
% % PACS codes here, in the form: \PACS code \sep code
% 
% \end{keyword}

%% ----- END ELSEVIER STUFF -----

\end{frontmatter}

%% ----- ICARUS STUFF -----
%% Some formatting on the first, second, and third pages are required
%% by the Icarus Editorial Office that are not required by Elsevier.
%% This section contains those things.  When you are ready to transition
%% to ``Elsevier'' mode, copy your abstract and keywords up into
%% the ELSEVIER STUFF section, and then you can just delete everything
%% in this section.

%% We need to list the number of manuscript pages, figures, and tables. 
%%
%% Rather than manually count these things out, we'll use a little
%% trick here from Paul.  All you have to do is place three \label{}
%% tags on the last page, the last table, and the last figure, that
%% way these values are automatically updated (as long as you remember
%% to move the lasttable and lastfig labels when you add or remove
%% tables and figures).

\begin{flushleft}
\vspace{1cm}
Number of pages: \pageref{lastpage} \\
Number of tables: \ref{lasttable}\\
Number of figures: \ref{lastfig}\\
\end{flushleft}

%% Don't worry about finding the various last* tags and deleting them
%% when you go to ``Elsevier'' mode if you don't want to, they should be
%% silently ignored.

%% The second page should indicate a proposed running head of not more 
%% than 55 characters, and the name and address to which editorial 
%% correspondence and proofs should be directed.  The pagetwo 
%% environment that icarus.sty provides will make page two for you,
%% just give the running head as an argument to the environment, and
%% then your correspondence address inside.
\begin{pagetwo}{Colors of asteroid pairs}
%                        1         2         3         4         5
%                           1234567890123456789012345678901234567890123456789012345

Nicholas A. Moskovitz\\
Department of Terrestrial Magnetism\\
Carnegie Institution of Washington\\
5241 Broad Branch Road\\
Washington, DC 20008, USA. \\
\\
Email: nmoskovitz@dtm.ciw.edu\\
Phone: (202) 478-8862 \\

\end{pagetwo}

%% start numbering lines
 \linenumbers

\begin{abstract}

Recent dynamical studies have identified pairs of asteroids that reside in nearly identical heliocentric orbits. Possible formation scenarios for these systems include dissociation of binary asteroids, collisional disruption of a single parent body, or spin-up and rotational fission of a rubble-pile. Aside from detailed dynamical analyses and measurement of rotational light curves, little work has been done to investigate the colors or spectra of these unusual objects. A photometric and spectroscopic survey was conducted to determine the reflectance properties of asteroid pairs. New observations were obtained for a total of 34 individual asteroids. Additional photometric measurements were retrieved from the Sloan Digital Sky Survey Moving Object Catalog. Colors or spectra for a total of 42 pair components are presented here. The main findings of this work are: (1) the components in the observed pair systems have the same colors within the uncertainties of this survey, and (2) the color distribution of asteroid pairs appears indistinguishable from that of all Main Belt asteroids. These findings support a scenario of pair formation from a common progenitor and suggest that pair formation is likely a compositionally independent process. In agreement with previous studies, this is most consistent with an origin via binary disruption and/or rotational fission.

\end{abstract}

% %% Keywords should appear after the abstract. 

%% ----- END ICARUS STUFF -----

%main text
\section{Introduction \label{sec.intro}}

Analyses of osculating orbital elements of Main Belt asteroids have revealed over 80 pairs of asteroids that reside in nearly identical heliocentric orbits \citep{Vok08,Pravec09,Rozek11}. These objects are distinct from binary asteroids as they are not on bound orbits around a common center of mass, and it is unlikely that their proximity is due to random fluctuations of asteroid densities in orbital element space. Backwards integration of these pairs' heliocentric orbits suggests they may have separated recently into an unbound state, in some cases much less than a Myr ago \citep{Vok09,Vok09b,Pravec10,Vok11,Duddy12}. As such these are interesting objects for studying phenomena, such as space weathering and radiation pressure forces, that are relevant to the ongoing dynamical, physical and chemical evolution of Main Belt asteroids.

The components of known pairs are typically a few km in size and consist of a primary and a secondary (respectively defined as the larger and smaller components based on measured absolute magnitudes). One formation scenario for these systems \citep{Scheeres07,Pravec10} involves parent asteroids that were spun up to a critical frequency by the YORP effect, i.e. a change in angular momentum due to anisotropic emission of thermal photons \citep{Rubincam00,Bottke06}. At this critical frequency the parent would fission into a proto-binary system and eventually disrupt under its own internal dynamics to form an unbound asteroid pair \citep{Jacobson11}. The estimated size ratios and observed rotational properties of known pair systems are consistent with a formation scenario via rotational fission \citep{Pravec10}. Progressive mass shedding due to YORP spin-up and accretion of a dynamically unstable proto-satellite offers a similar pathway to pair formation \citep{Walsh08}. The close spectral similarity between components in one pair system \citep{Duddy12} and the photometric similarity between components in another \citep{Willman10} support these scenarios.

Though a fission origin is consistent with the size ratios and rotation properties for a large number of pairs \citep{Pravec10}, collisions provide another possible formation mechanism that may explain a subset of systems. In this scenario a catastrophic collision would produce a distribution of fragments, of which only the largest two are observed as an associated pair. Collisions between small bodies can result in compositionally complex outcomes \citep[e.g.][]{Leinhardt09}, but it is unclear how collisional formation would affect the relative colors and/or spectra of the km-scale objects found in pair systems. Aside from any spectroscopic or photometric implications, hydrodynamic simulations of impact events make predictions about the resulting orbital properties and size ratios of collisional fragments \citep{Nesvorny06,Durda07}. Unfortunately, due to incompleteness for sub-km bodies in the Main Belt, it is currently not possible to  fully test these predictions \citep{Vok08}.

A final formation mechanism involves the dynamical dissociation of bound binary systems. Perhaps the best evidence for ongoing pair formation via binary disruption comes from the system of asteroids associated with 3749 Balam. Adaptive optics and light curve observations \citep{Merline02,Marchis08a,Marchis08b,Polishook11} have shown that  Balam has two bound satellites, making it a rare triple system. One of its two satellites is on a highly eccentric orbit ($e\sim0.9$), while the other orbits at a distance of only 20 km (Balam itself is about 5-10 km in size). In addition, this triple system has a dynamically associated pair, asteroid (312497) 2009 BR60. Backwards numerical integrations show that the orbits of Balam and 312497 converge within the past 0.5 Myr \citep{Vok09b}. Numerical models suggest that a cascade of fragments, like that seen in the Balam system, can result from repeated rotational fission events \citep{Scheeres11}. It seems in this case that YORP fission and binary dissociation may be closely related processes.

However, it is unclear how the compositions of asteroid pairs would reflect an origin due to the dissociation of binaries. Component-resolved spectra have been obtained for a very small number of bound binary systems \citep[e.g.][]{Polishook09,Marchis11,DeMeo11}. This is in part due to the need for either adaptive optics or space based observations to resolve the individual components. Available spectra suggest that the components of binaries are compositionally similar, however data is scarce and more information is required before generalized statements can be made.

In light of these various formation mechanisms, there remain several unaddressed questions regarding the formation and evolution of asteroid pairs. For instance, the relationship between pairs and bound multi-component systems is unclear. In addition, little is known about their compositions/taxonomic types. Different formation mechanisms could affect the relative compositions of pair components in different ways. However, for each of these mechanisms the physics of separation should predominantly depend on the internal (rubble pile) structure and density of the parent asteroid. Objects in the size range of asteroid pairs are expected to be rubble piles \citep{Pravec00}. It is possible that a variety of formation mechanisms are responsible for the formation of the ensemble population of asteroid pairs. Reflectance spectroscopy or photometric colors may help to provide arguments regarding the mechanism of formation on a case-by-case basis.

Here we present a survey of asteroid pairs to constrain their spectro-photometric properties. The observations, data reduction and error analysis are presented in \S2. This data set provides the means to investigate the relative reflectance properties of pair primaries and secondaries, and facilitates a direct comparison to the color distribution of ordinary Main Belt asteroids (\S3). The results and implications of this survey are discussed in \S4.

\section{Observations and Data Reduction \label{sec.obs}}

New optical observations of 34 individual asteroids were performed. In the majority of cases {\it BVRI} photometric colors were measured. Visible wavelength spectra were obtained for two objects, and for one object Gunn {\it gri} photometry was obtained. In addition, {\it ugriz} photometry of 16 pair components were retrieved from the 4th release of the Sloan Digital Sky Survey Moving Object Catalog \citep[SDSS MOC, ][]{Ivezic01}. Six of these SDSS pairs overlap with our sample. In total, data were acquired for at least one component in 30 pair systems and for both components in 12 pair systems. A previously reported pair that has since been found to be spurious was also observed (see \S\ref{subsec.compare}). Targets for observation were selected from \citet{Pravec09} based on observability during the scheduled observing runs. Table \ref{tab.obs} summarizes the observing circumstances for each target.

\subsection{Photometry \label{subsec.phot}}

Photometric observations were conducted at three different facilities: with IMACS at the Magellan Baade 6.5m telescope at Las Campanas Observatory in Chile, with the SITe2k CCD at the DuPont 2.5m telescope also at Las Campanas, and with the SNIFS instrument \citep{Lantz04} at the University of Hawaii 2.2m telescope on Mauna Kea. IMACS was operated using the $f$/2 camera, which has a 27.5' field-of-view covered by a mosaic of eight 2k x 4k CCDs with plate scales of 0.2 "/pixel. The SITe2k at DuPont is a 2k x 2k CCD with an 8.85' field-of-view with a plate scale of 0.259 "/pixel. SNIFS was operated in imaging mode with its 4k x 4k CCD covering a field-of-view of approximately 9'  at a plate scale of 0.137 "/pixel. Observations were performed in February and August of 2010 with the SITe2k at DuPont, in March and August of 2010 and in March of 2011 with IMACS at Magellan, and in February of 2012 with SNIFS at the UH2.2m. The telescopes were operated in non-sidereal tracking modes when integration times were long enough to have resulted in trailing greater than the measured seeing. When using on-chip calibration stars (see below) we tracked at 1/2 the non-sidereal rate of the asteroid such that the asteroid and field stars shared a common point spread function. When not using on-chip calibration stars tracking was performed at the non-sidereal rate of the asteroid.

Different filter sets were used at each facility. With IMACS Bessell {\it BVR} and CTIO {\it I} filters were used. The band centers of these filters are 0.44, 0.55, 0.64, and 0.82 $\mu m$ respectively. At the DuPont telescope, Johnson {\it BV} and Kron-Cousins $R_cI_c$ were employed. The  band centers for these filters are the same as the Bessell set with the exception of $I_c$ with a center at 0.85 $\mu m$. With SNIFS the Gunn {\it gri} filter set was used with band centers of 0.47, 0.62 and 0.75 $\mu m$. Data retrieved from the SDSS MOC correspond to {\it ugriz} filters centered at 0.35, 0.47, 0.62, 0.75, and 0.89 $\mu m$ respectively. 

Observations were typically conducted in sequences with interspersed $V$- or $r$-band measurements to monitor variability due to observing conditions or rotation of the asteroid. For example, a single observing sequence would involve taking a set of seven images with a filter order of $VBVRVIV$. In this case the mean value of adjacent $V$-band magnitudes would be used to compute the $B-V$, $V-R$ and $V-I$ colors. In some cases only single exposures were taken in each band. Though this was not ideal for producing perfectly calibrated photometry, the gain in observing efficiency allowed us to increase our sample to a statistically significant size. Any errors introduced by taking single exposures (for instance not being able to correct for light curve variability) should be random and small since fewer than 20 minutes were typically needed for single $BVRI$ exposures.

Data reduction employed standard IRAF routines for overscan correction, bias subtraction, flat fielding, and aperture photometry. Aperture radii were set to minimize the photometric error of the asteroid measurement. Since, the asteroids were always fainter than or roughly equal in brightness to the calibration stars, the errors introduced by background noise from the asteroid measurement always dominated the net photometric uncertainty. Choosing an aperture that minimized photometric noise ensured the highest quality results. With median seeing around 0.65" at Las Campanas, typical aperture radii of $\sim1-2"$ were employed. Background annuli were defined to have a radius of 4"  from the photocenter of the target and a width of 8". 

Photometric calibration was achieved in one of two ways. When the target was in an SDSS field, up to 15 on-chip stars with solar-like colors were used to calibrate the instrumental magnitude of the asteroid by determining a single photometric correction factor based on the difference between the SDSS stars' measured and catalog magnitudes. Definitions of solar colors from the SDSS website\footnote{http://www.sdss.org/dr4/algorithms/sdssUBVRITransform.html} and {\it BVRI} $\rightarrow$ {\it ugriz} transforms \citep{Jester05} were employed. SDSS stars much more than two or three magnitudes brighter than the asteroid were often saturated, while those much fainter than the asteroid were not used to avoid introducing extra error into the photometric calibration. This resulted in using field stars with magnitudes roughly comparable to or a few magnitudes brighter than the asteroid. When SDSS field stars were not available, Landolt standard fields were observed at a range of airmasses and at several times throughout the night to fit a photometric transform solution (zero point, airmass term and color term). As a test, several objects were calibrated with both techniques. In all tests the resulting photometry was consistent within 0.1 magnitudes. Apertures for calibration stars (both Landolt and SDSS) were set equal to that used for each asteroid.

Tables \ref{tab.bvri} and \ref{tab.ugriz} summarize the results of our collected photometry. Phase corrected absolute magnitudes ($H_V$) are calculated for each object in Table \ref{tab.bvri}. Phase and distance corrections to the measured apparent magnitudes were applied using the observational data in Table \ref{tab.obs} and the formalism of the IAU H-G system \citep{Bowell89}. A slope parameter of 0.15 was assumed for all asteroids. No attempts were made to accurately account for the full rotational variability of the targets. Accounting for the possible range of slope parameters and light curve amplitudes suggests that the calculated absolute magnitudes  are accurate to approximately $\pm$0.5. In general these absolute magnitudes are several tenths higher than those reported by the Minor Planet Center and JPL Horizons. This is a well known bias in these catalogs that may be attributed to the use of open, unfiltered images for the majority of reported measurements \citep{Juric02}.

Tables \ref{tab.bvri} and \ref{tab.ugriz} make several notes about individual objects. It is noted when an object was observed on several occasions, either with different instruments or by SDSS. It is noted when multiple {\it BVRI} or {\it gri} sequences were obtained for a given object within a night. The magnitudes and colors for objects with multiple sequences represent mean values weighted by the signal-to-noise ratios of the individual observations. A full listing of all $V-$band magnitudes from individual exposures are included in the online Supplementary Data. The error bars in Tables \ref{tab.bvri} and \ref{tab.ugriz} are addressed in \S\ref{subsec.error}. Finally, it is noted when SDSS field stars were used for photometric calibration. All other observations were calibrated with Landolt standards.
	
The data for one complete pair (asteroids 17288 and 203489) are shown in Figure \ref{fig.colorcomp}. These data have been solar-corrected assuming solar colors (from the SDSS website) of $B-V$ = 0.65, $V-R_C$ = 0.36, and $R_C-I_C$ = 0.32, and thus provide coarsely sampled reflectance spectra of these asteroids. All subsequent presentations of our data will show solar-corrected values. This is not the case for the photometry in Tables \ref{tab.bvri} and \ref{tab.ugriz}.

\subsection{Spectroscopy \label{subsec.phot}}

In two cases the targets were sufficiently bright that visible wavelength (0.45-0.82 $\mu m$) spectra were obtained with IMACS operating in its long-slit, low-resolution mode with a 200 lines/mm grism. These settings produce a single-order spectrum that spans two of the IMACS 2k x 4k chips at a dispersion of approximately 2 \AA/pixel with a small gap in wavelength coverage around 0.6 $\mu m$. A blocking filter with a cutoff at 0.455 $\mu m$ was employed to prevent contamination from higher orders. The observing circumstances for the two spectroscopic targets are summarized in Table \ref{tab.spec}.

Reduction of these spectra employed standard IRAF and IDL routines to overscan correct, bias subtract, flatten, extract, dispersion correct, combine, normalize and re-bin the data. Solar analog stars were observed close in time and pointing location to the asteroids (Table \ref{tab.spec}). In the case of asteroid 10123 and its analog SA104-335, both spectra were obtained within a span of 30 minutes and at a difference in airmass of 0.04. For asteroid 99052 and its analog HD127913, the two spectra were obtained within 40 minutes of one another and at a difference in airmass of 0.03. The angular separations of the asteroids and their respective analogs were less than $20^\circ$. The extracted asteroid spectra were divided by their respective analogs and normalized to produce relative reflectance spectra. Combined He, Ne and Ar arc lamp spectra were obtained immediately after each target to provide pointing-specific dispersion solutions.

Figure \ref{fig.99052} presents both a spectrum and photometry for asteroid 99052. These two techniques produce similar spectral profiles with only slight differences in slope. Light curve variability was not accounted for in the photometric data and thus could be the cause of its slightly redder slope, however the light curve period of 99052 is currently unknown. Another factor could be that the spectroscopic calibration star HD127913 (Table \ref{tab.spec}) is not a perfect solar analog. 

The observed spectra provide the means for more accurate taxonomic classification than can be achieved with broad band photometry. We assign a taxonomic type to each of the observed spectra by conducting a chi-squared minimization search through the mean spectral values for each of the SMASS taxa \citep{Bus02}. The best-fit taxa are given in Table \ref{tab.spec}.

\subsection{Consistency Checks and Error Analysis \label{subsec.error}}

\begin{figure}[]
\begin{center}
\includegraphics[width=14cm]{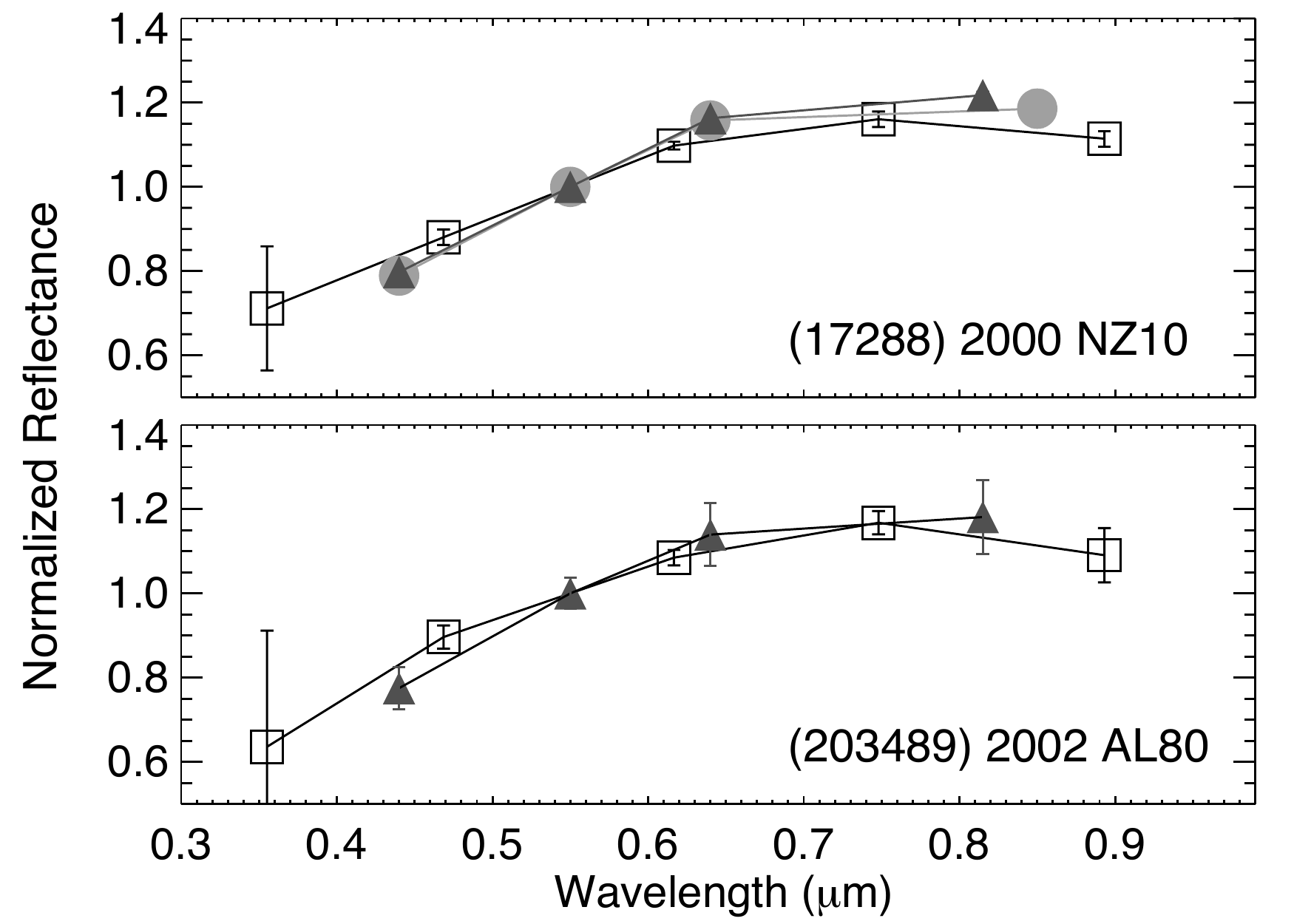}
\end{center}
\caption[Asteroid Pair 17288-203489]{Spectro-photometry of asteroid pair primary 17288 (top) and secondary 203489 (bottom). SDSS {\it ugriz} photometry is shown as open squares. {\it BVRI} photometry from IMACS/Magellan is shown as dark grey triangles and from DuPont as light grey circles.  Some error bars fall within the size of the symbols. The data are normalized at 0.55 microns. The repeat observations of each object are highly consistent. The reflectance profiles of these two objects are very similar.
} 
\label{fig.colorcomp}
\end{figure}

\begin{figure}[]
\begin{center}
\includegraphics[width=14cm]{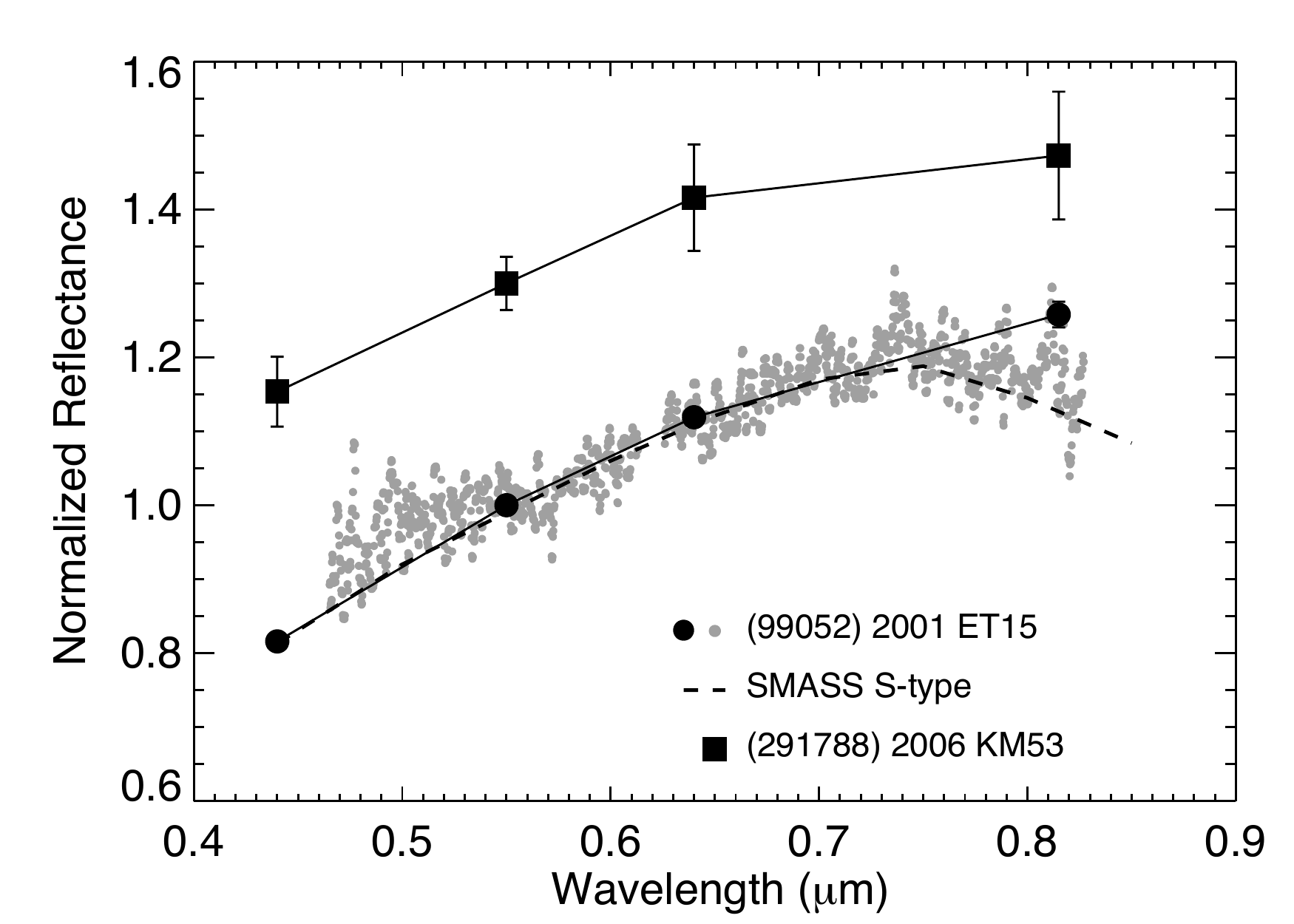}
\end{center}
\caption[Asteroid Pair 99052-291788]{Reflectance measurements of pair 99052-291788. {\it BVRI} photometry and a spectrum were measured for the primary 99052. The two different techniques produced very similar results. The spectrum is best fit by an S-type in the SMASS taxonomic system (dashed). The data for the secondary 291788 are vertically offset by 0.3 units and closely match the spectro-photometry of 99052.
} 
\label{fig.99052}
\end{figure}

Our data set provides several methods for checking the quality and repeatability of the observations. The most obvious check is to compare intra- and inter-night observations of the same object. Repeat intra-night observations of the same object generally resulted in less than 0.05 magnitude variability in colors. As such the colors in Tables \ref{tab.bvri} and \ref{tab.ugriz} represent weighted means of all measured colors within a night. Asteroid 17288 was observed with IMACS/Magellan and the SITe2k/DuPont on two separate nights. The colors from these observations are very similar (Fig. \ref{fig.colorcomp}). Asteroid 143662 was observed on two separate nights with the SITe2k/DuPont. The photometry for this object shows some inconsistency with colors differing up to $0.07$ magnitudes between nights but are still consistent within the errors (Table \ref{tab.bvri}). 

Asteroid 195479 was observed with IMACS/Magellan and the SITe2k/DuPont on two separate nights. The photometry roughly agrees between these sets of observations, however the $V-R$ colors differ by 0.08 magnitudes (Table \ref{tab.bvri}). This could be due to the use of different calibration methods, i.e. using SDSS field stars versus Landolt standards. Another possible cause is the lack of any light curve corrections to the IMACS data. In the case of the IMACS observations, a single $BVRI$ sequence was obtained, taking less than 25 minutes to complete. With the DuPont data, light curve corrections were possible and applied because the brightness of 195479 monotonically increased by two-tenths of a magnitude over the 2.5 hours necessary to complete the exposure sequence. This  variability was much larger than the photometric errors and thus a strong indication of light curve variability. Regardless of the cause of this offset in $V-R$ color, this is the largest discrepancy amongst any of our repeat observations.

Another consistency check is to compare our data to that of objects included in the SDSS MOC. This overlap includes 6 objects (Table \ref{tab.ugriz}) and in all cases the difference in colors is no more than 0.06 magnitudes. For 4 of these 6 the color differences are less than 0.02 magnitudes. Figure \ref{fig.colorcomp} shows the close match between SDSS photometry and new observations for two asteroids. The close match between our spectroscopic and photometric observations of 99052 (Fig. \ref{fig.99052}) provides a final confirmation of data quality.

Based on the general consistency of these data, a {\it conservative} estimate to the systematic uncertainties in the photometric calibration is $\pm0.1$ magnitudes. This estimate accounts for issues of uncorrected light curve variability, non-photometric observing conditions, and the use of imperfect calibration stars and is unrelated to the signal-to-noise of the observations. We adopt this limit as an estimate to the overall accuracy of this survey and employ it in a statistical comparison of pair colors (\S\ref{sec.analysis}). This limit to the systematic uncertainties is not large enough to affect any of the following conclusions or analysis.

The error bars reported in Table \ref{tab.bvri} merit further discussion. Due to the large range of observed magnitudes (from 17.5 up to 22.5), the measurement uncertainty on the brightest targets will be dominated by systematic errors, whereas background noise dominates for the fainter targets. It is unclear where the transition between these two regimes exists and, as the previous discussion indicates, it is non-trivial to estimate systematic errors. Therefore the errors reported here fall into one of two categories. In the case of objects observed with a single {\it BVRI} sequence, the errors represent the signal-to-noise of the data as set by background noise levels with standard error propagation for the colors. In the case of objects observed with multiple {\it BVRI} sequences (as denoted by $^c$ in Table \ref{tab.bvri}), the uncertainty on the V-band magnitudes represents one standard deviation across all measurements. Hence, the standard deviations of these V-band magnitudes provide a lower limit to the light curve amplitude of the targets even though no full rotational light curves were resolved. The best example of a partially resolved light curve was for asteroid (69142) 2003 FL115, which monotonically increased in brightness from $V=20.3$ to $V=19.9$ across 10 exposures during the hour and a half in which it was observed. Hence the $\pm0.19$ uncertainty on the calibrated V-band magnitude for 69142 in Table \ref{tab.bvri} represents the standard deviation of these 10 measurements and is likely indicative of light curve variability of the asteroid.    The uncertainties on the colors for these multi-sequence targets represent the errors on the weighted means.

\section{Analysis \label{sec.analysis}}

Comparison of multi-band photometry is facilitated by assigning a single diagnostic parameter to each set of multi-band measurements. An obvious choice for this parameter is a modification for the {\it BVRI} filter set of the $a^*$ principal component color defined by \citet{Ivezic01}:
\begin{equation}
a^* = 0.908\cdot(B - V) + 0.409\cdot(R_C - I_C) - 0.856.
\end{equation}
This modification is based on the  {\it BVRI} $\rightarrow$ {\it ugriz} transform equations of \citet{Jester05} and redefines the  $B-V$ versus $R_c - I_c$ color space to maximize the separation of C- and S- complex asteroids, thus providing a means for coarse taxonomic assignment. The utility of $a^*$ as a tool for such taxonomic assignment is due to its strong correlation with spectral slope \citep{Nesvorny05}. This principal component color is computed for each of the asteroids in Tables \ref{tab.bvri} and \ref{tab.ugriz}. The errors on $a^*$ represent a propagation of the reported $B-V$ and $R_C-I_C$ errors through Equation (1).

\subsection{Comparison of Primary and Secondary Colors \label{subsec.compare}}

We first compare $a^*$ colors of the primary and secondary components (Table \ref{tab.comp}, Fig. \ref{fig.pca}). The $a^*$ colors of the components in each pair are the same within the reported error bars. Table \ref{tab.comp} includes an assignment of taxonomic complex for each of the pairs. Only the S- and C-complexes are considered here and are distinguished by positive or negative $a^*$ values respectively. We make no attempt to distinguish specific taxonomic types due to the coarse spectral sampling of the broad band photometric filters. In reality this rough taxonomic assignment lumps a variety of classes into each complex. Asteroids with positive $a^*$ include the S-, D-, A-, R-, L- and V-classes, while those with negative $a^*$ include C-, B- and X-types \citep{Ivezic01,Bus02}. 

Figure  \ref{fig.pca} presents observations for the 11 complete pairs; no SDSS or spectroscopic data are included. The components for a majority (10/11) of these pairs have the same a* colors within the estimated $\pm0.1$ magnitude systematic uncertainties. The pair 69142-127502 falls outside of the limits set by the estimated uncertainties. The spectro-photometry for this pair is shown in Figure \ref{fig.69142}. The $a^*$ values for these asteroids are $0.15\pm0.12$ and $0.04\pm0.14$ respectively. These colors are indistinguishable within the error bars, which are some of the largest reported here (Tables \ref{tab.bvri} and \ref{tab.ugriz}). The photometric similarity of these objects is apparent in Figure \ref{fig.69142}. The large difference in their $a^*$ colors is most likely a consequence of the different filter sets and highlights the limitations of comparing individual objects (as opposed to a statistical comparison) based on broad band photometric measurements.

The pair 10123-117306 is not included in Figure \ref{fig.pca}. This is due to an incomplete {\it BVRI} sequence for 117306 caused by the asteroid passing over a field star during the {\it I}-band exposure. This made it impossible to reliably calibrate the photometry in that band and to calculate an $a^*$ value. Nevertheless, the spectral profiles of the asteroids in this pair, particularly the unusually red slope, are very similar (Fig. \ref{fig.10123}). It is reasonable to suggest that the reflectance properties of 10123-117306 would result in this pair plotting near the slope-one line in Figure \ref{fig.pca}.

Data were also obtained for the reported pair 34380-216177 (Fig. \ref{fig.34380}). The photometry for this system resulted in significantly different $a^*$ colors for the two components (Tables \ref{tab.bvri} and \ref{tab.ugriz}). This does not have a bearing on our results: the probability of association for this pair was one of the lowest in \citet{Pravec09} and updated dynamical integrations failed to reveal an orbital convergence within the past 2 Myr (P. Pravec, private communication, April 2012). Therefore, this pair is now considered spurious and is not included in the analysis here.

The significance of the apparent correlation between pair colors in Figure \ref{fig.pca} is tested by comparing $a^*$ colors for random asteroids selected from the SDSS MOC. To perform this test $a^*$ is calculated for a randomly selected Main Belt asteroid. An $a^*$ color is then calculated for the next closest object in orbital element space as defined by the distance metric of \citet{Zappala94}. This pseudo-pair is plotted on the axes of Figure \ref{fig.pca} and the process is repeated 12 times to mimic the set of complete pair systems (including 10123-117306) observed here. The number of pseudo-pairs that fall outside of the $\pm0.1$ region is then recorded. This Monte Carlo style selection of 12 pseudo-pairs is repeated 10,000 times. This test shows that in only 1.7\% of the trials do 11 out of the 12 psudeo-pairs have $a^*$ colors within $\pm0.1$ of each other. In other words, 98.3\% of these trials show a distribution of colors less correlated than those in Figure \ref{fig.pca}. Only 0.2\% of the time do all 12 pseudo-pairs have $a^*$ colors within $\pm0.1$ of each other. This would be the case if we assume that additional data of pair 69142-127502 revealed a closer match in $a^*$. This test suggests with greater than 98\% significance that the correlation between $a^*$ colors for asteroids pairs is not a random result. Reducing the conservative estimate for the systematic uncertainties of this survey in half to $\pm0.05$ increases the probability of a non-random $a^*$ distribution to $>99$\%, further supporting the conclusion that the components in asteroid pair systems have the same colors. 

\subsection{Comparison of Pairs to Main Belt Asteroids \label{subsec.compare}}

The $a^*$ colors of all 40 asteroid pairs presented here (our 30 complete photometric observations plus 10 from SDSS and excluding the spurious pair 34380-216177) are compared to those of Main Belt asteroids from the SDSS MOC (Fig. \ref{fig.histcomp}). The bi-modality in these histograms is indicative of the $a^*$ color separation of C- and S-complex asteroids. From this figure it is clear that, by definition, an $a^*$ color cut at 0.0 roughly segregates these two complexes (Table \ref{tab.comp}). However, we note that $a^*$ is not meant to provide detailed taxonomy and that the assignment of a taxonomic complex to objects with error bars that straddle the 0.0 cut is somewhat arbitrary due to the the clear overlap of the C- and S-complexes at these values. This ambiguity in taxonomic typing is largely irrelevant to this study, but could be resolved with future spectroscopic observations.

Both histograms in Figure \ref{fig.histcomp} are influenced by observational biases. This is clear in the apparent over-abundance of S-types. In fact debiased estimates suggest that C-types are more numerous by a factor of more than two to one in the Main Belt \citep{Ivezic01}. For our observations, no attempts were made to obtain an unbiased photometric sample of asteroid pairs. In spite of these limitations, it is reasonable to suggest that to first order the biases inherent to SDSS also affected our observations. We were magnitude limited to those objects that could be observed on a specific date, in similar fashion to detection limits inherent to the SDSS. Therefore a qualitative comparison of these histograms is justified.

Though both histograms in Figure \ref{fig.histcomp} have bi-modal distributions, it appears that the C-type pairs are  slightly shifted towards positive $a^*$ colors. However, this is likely a result of low number statistics and not a robust trend. A two-sided Kolmogorov-Smirnov test suggests the likelihood that these distributions are drawn from different parent populations is significant at only the 1.4-sigma level. Hence, the colors of asteroid pairs observed in this survey are statistically indistinguishable from the overall population of Main Belt asteroids.

\section{Discussion \label{sec.disc}}

\begin{figure}[]
\begin{center}
\includegraphics[width=14cm]{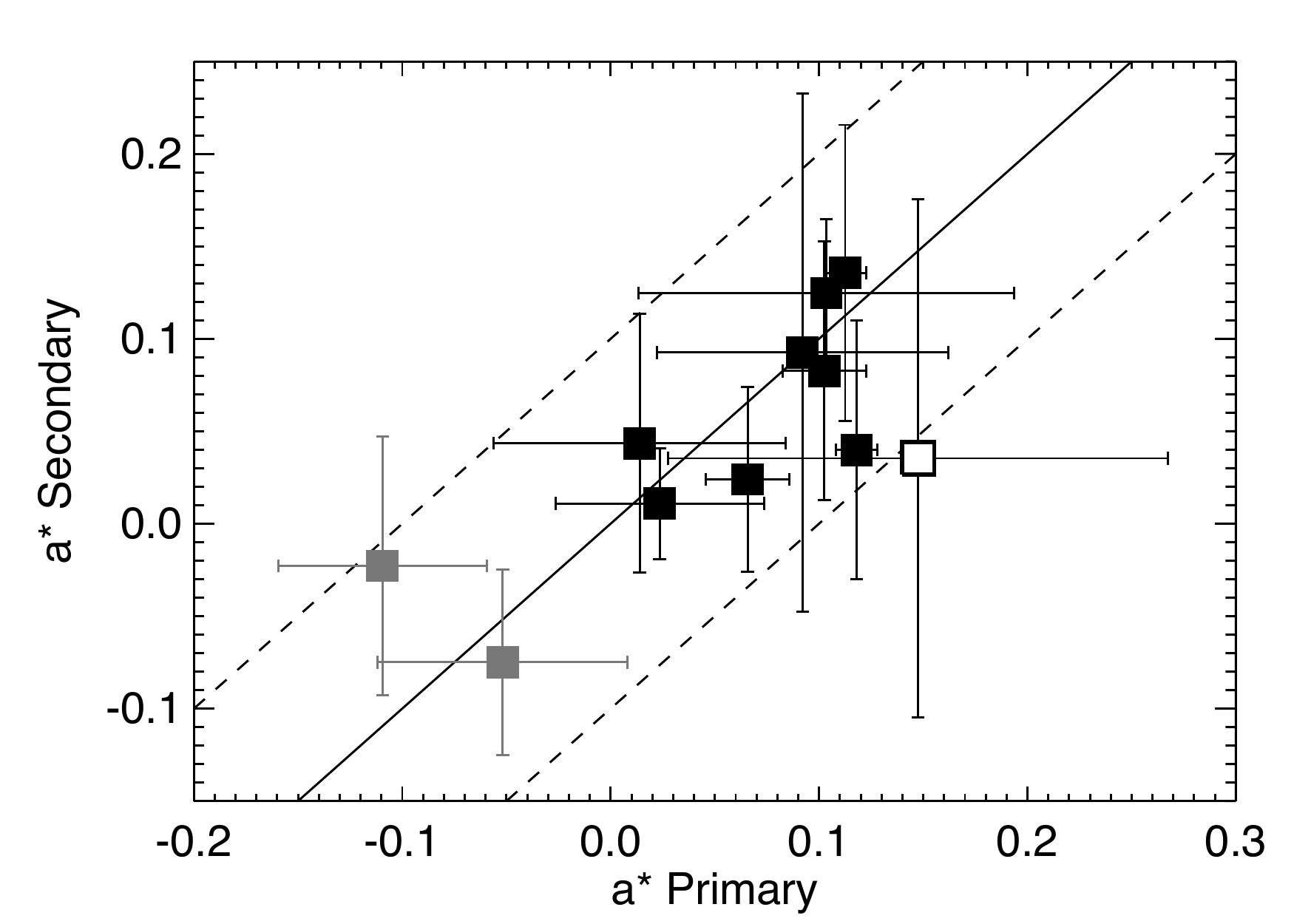}
\end{center}
\caption[a* Comparison]{Comparison of $a^*$ colors for the primary and secondary components in 11 pair systems. The solid line has a slope of one, representing a perfect match of primary and secondary colors. The dashed lines represent estimates on the systematic uncertainties ($\pm$0.1) of our observations. The grey symbols indicate $a^*$ colors consistent with the C- taxonomic complex ($a^*<0.0$), the black symbols indicate colors consistent with the S-complex ($a^*>0.0$). The open symbol is the only pair that falls outside of our systematic uncertainties. This distribution of colors is non-random at a significance level of $>$98\%.
} 
\label{fig.pca}
\end{figure}

We have presented results and analysis of a spectro-photometric survey of dynamically associated asteroid pairs. A combination of new observations and archival data from the SDSS MOC have provided insight on the reflectance properties of 44 individual asteroids in 30 pair systems and one spurious pair. Data were obtained for both components in 12 pair systems. These data suggest a correlation between the colors of primary and secondary components (Fig. \ref{fig.pca}) at greater than 98\% significance. We suggest this argues in favor of a common origin for these pairs. 

The components in one of the observed systems, 34380-216177, have significantly different colors (Fig. \ref{fig.34380}). However, updated dynamical integrations have revealed that this is a spurious pair (P. Pravec, private communication, April 2012) and thus should not have been included in our survey. This highlights the need for future follow-up observations of other dynamically identified pairs. 

\begin{figure}[]
\begin{center}
\includegraphics[width=14cm]{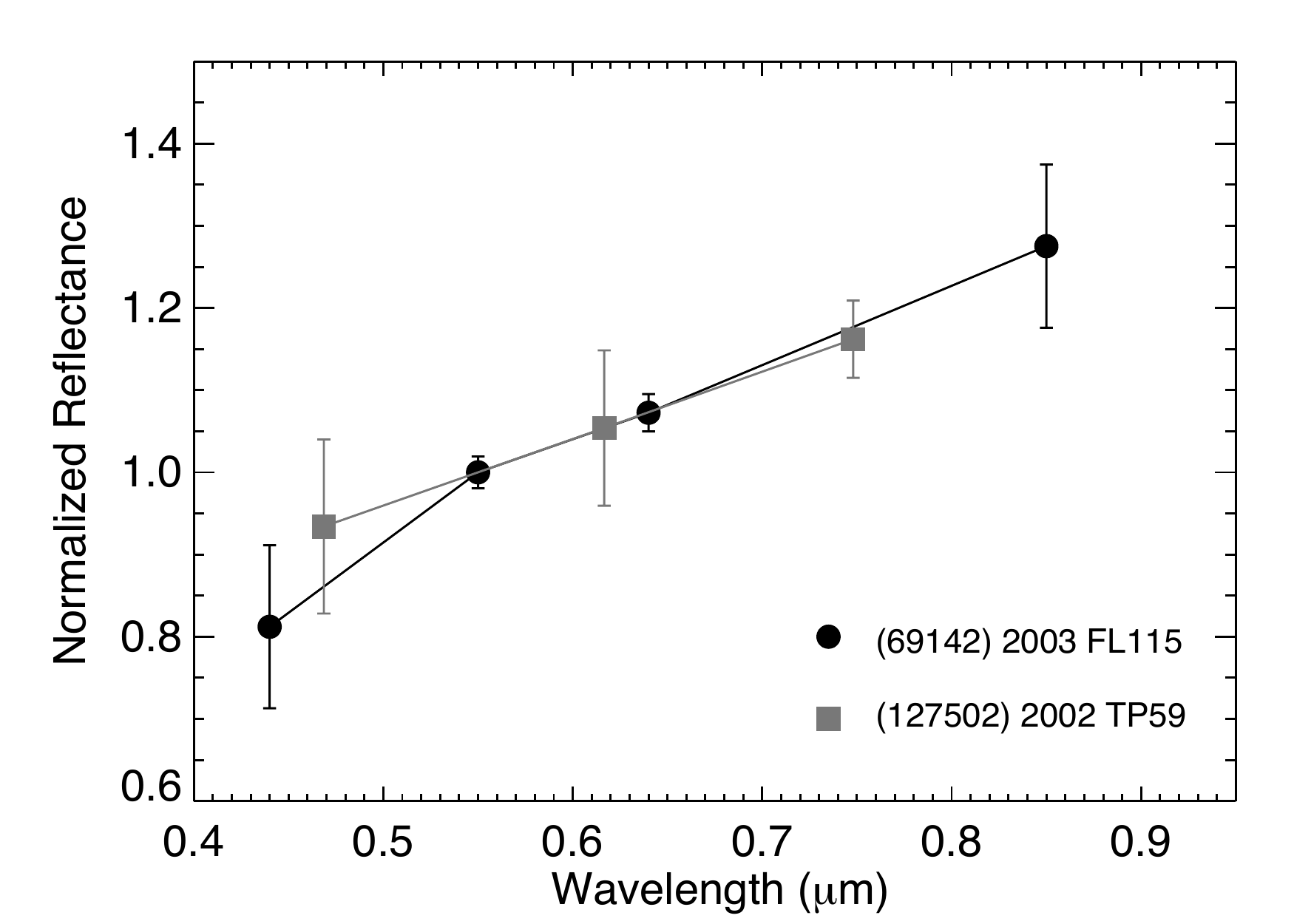}
\end{center}
\caption[Asteroid Pair 69142-127502]{Photometry of asteroid pair 69142 ({\it BVRI}) and 127502 ({\it gri}). Though the $a^*$ colors of these asteroids differ by more than 0.1 magnitudes, they are the same within the errors bars. 
} 
\label{fig.69142}
\end{figure}

\begin{figure}[]
\begin{center}
\includegraphics[width=14cm]{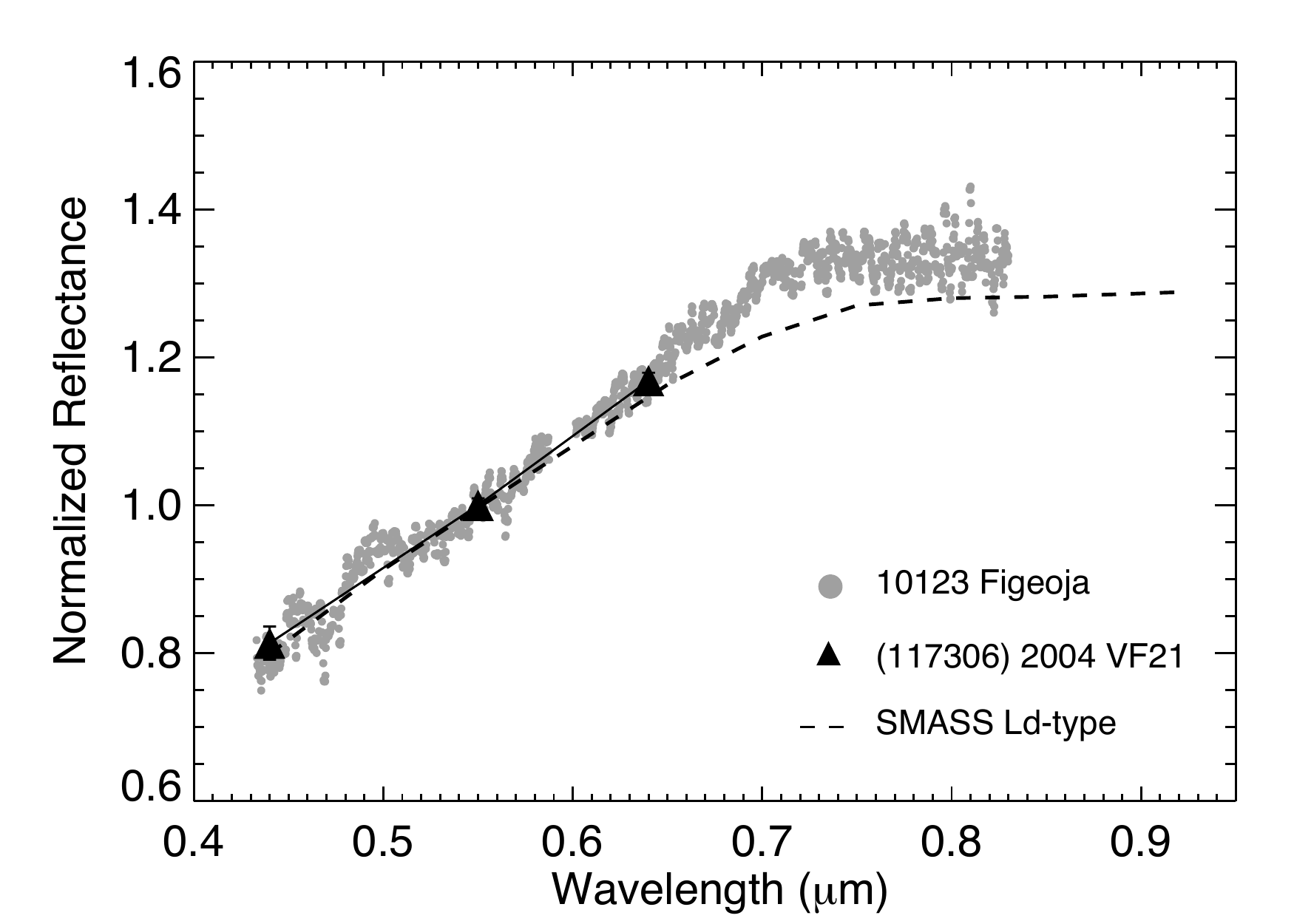}
\end{center}
\caption[Asteroid Pair 10123-117306]{Observations of asteroid pair 10123-117306. The IMACS spectrum of 10123 (grey dots) is indicative of an SMASS Ld-type (dashed line).  The photometric error bars for 117306 mostly fall within the size of the symbols. Only {\it BVR} images were obtained for the secondary 117306 because of interference by a background field star during the {\it I}-band exposure. Nevertheless the spectral profiles of these objects are very similar.
} 
\label{fig.10123}
\end{figure}

\begin{figure}[]
\begin{center}
\includegraphics[width=14cm]{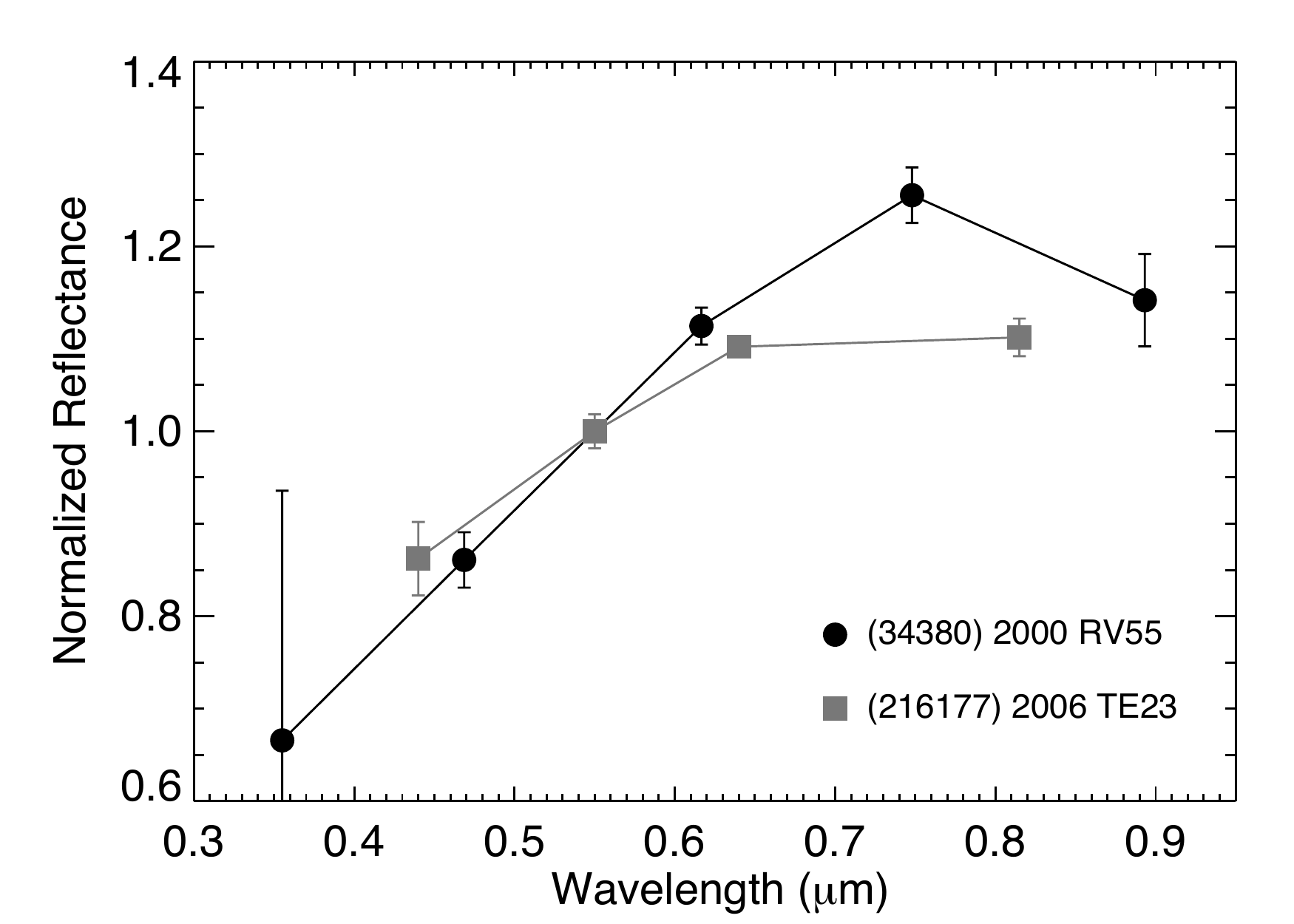}
\end{center}
\caption[Asteroid Pair 34380-216177]{Photometry of reported asteroid pair 34380 (SDSS {\it ugriz}) and 216177 ({\it BVRI}). Though originally identified as a pair with a low probability of association \citep{Pravec10}, updated dynamical integrations suggest that this pair is actually spurious.
} 
\label{fig.34380}
\end{figure}

Asteroids 69142-127502 are a second pair whose $a^*$ colors are not the same within the estimated $\pm0.1$ magnitude systematic uncertainties of this survey. However, within the rather large photometric errors for these objects, the spectral profiles and $a^*$ colors are the same (Fig. \ref{fig.69142}). The difference in filter sets used to observe this system may be the primary cause of its discrepant $a^*$ colors. Follow-up spectroscopy could confirm or refute any taxonomic or compositional link between these objects.

The results from several other studies can be used to compare pair reflectance properties. \citet{Duddy12} showed that the components in the pair 7343-154634 have very similar reflectance spectra. Data from \citet{Masiero11} show that the albedos of the pair 38395-141513, as determined by observations from the Wide-field Infrared Survey Explorer mission, are nearly indistinguishable with values of 0.0638 and 0.0623 respectively. This pair was included in our sample and has $a^*$ colors that differ by less than 0.03 magnitudes. These additional results support a compositional link between components and thus a common origin for pair systems.

Further comparison can be made to the survey of \citet{Ye11}. As part of a larger sample they observed 12 asteroids in 10 pair systems with data collected for two complete pairs (1979-13732 and 11842-228747). Unfortunately the data for one component in each of the completed systems were unreliable due to instrumental problems in one case and proximity to a bright field star in the second. As such it is difficult to draw conclusions regarding the relative colors of the components in these two systems. Four of the asteroids discussed here (2110, 4765, 15107, and 54041) were also part of the \citet{Ye11} survey. With two exceptions the data agree within the error bars. The $V-I$ colors for 15107 are significantly different: we measured $V-I=0.77 \pm 0.06$ whereas \citet{Ye11} measured $V-I=1.016 \pm 0.021$. The cause of this offset is not clear, but we note that our measured $a^*$ colors for 15107 and its companion 291188 are identical. The second discrepancy is for asteroid 4765: the data from \citet{Ye11}  suggest $a^*=0.06$ while SDSS MOC data suggest $a^*=-0.07$. Follow-up observations would help to clarify this inconsistency.

\begin{figure}[]
\begin{center}
\includegraphics[width=14cm]{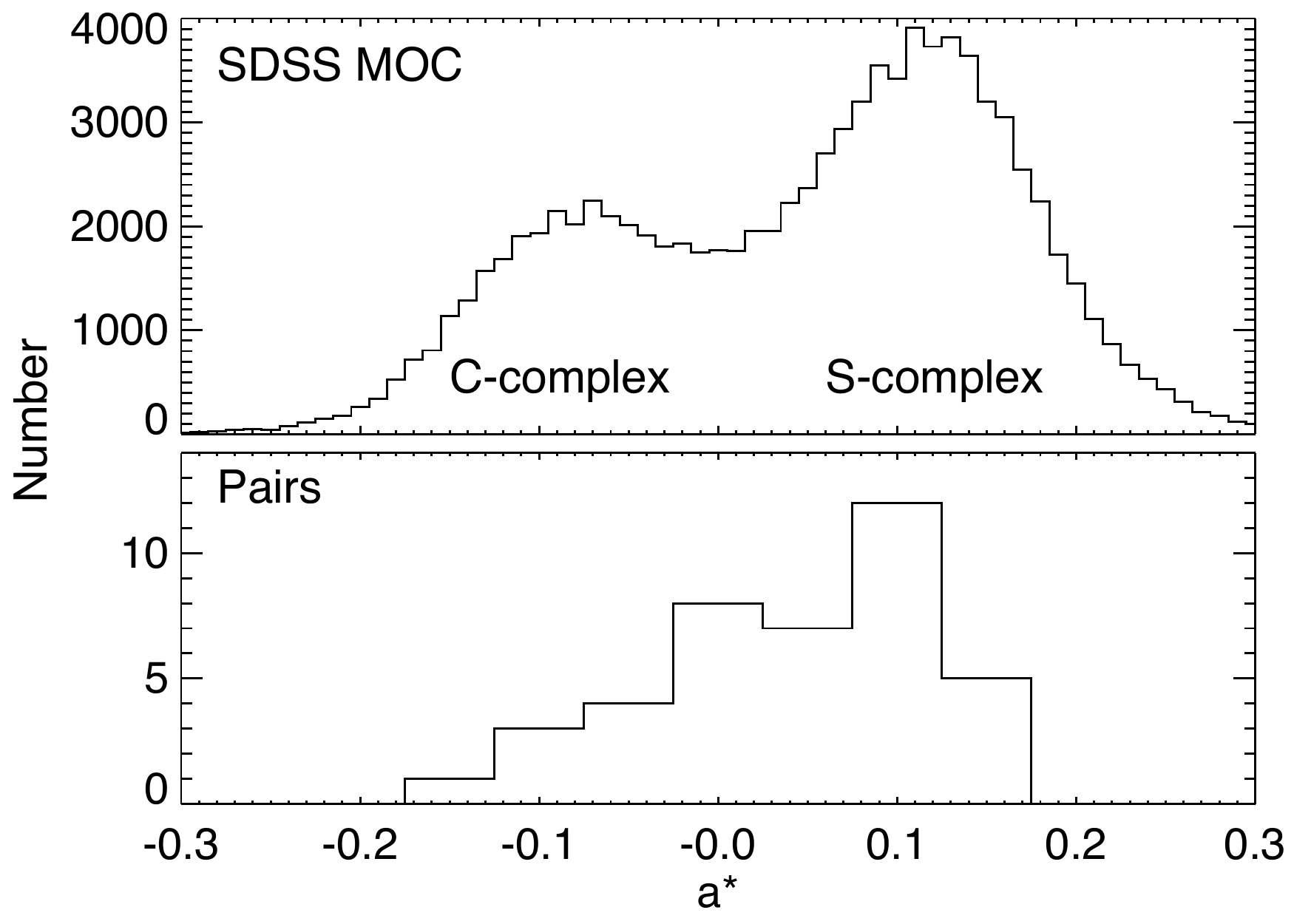}
\end{center}
\caption[a* Histograms]{Histograms of a* values for all objects within the SDSS MOC (top) and for all pairs considered in this study (bottom). The apparent bi-modality is attributed to the color difference between C- and S-complex asteroids. There is no statistically significant difference between these two distributions, suggesting that pairs represent an unbiased cross section of the compositional diversity of asteroids within the Main Belt.
} 
\label{fig.histcomp}
\label{lastfig}
\end{figure}

We have also shown that the $a^*$ distribution of pairs is similar to that of all Main Belt asteroids (Fig. \ref{fig.histcomp}). There appears to be no bias towards a single taxonomic complex. This strongly suggests that formation of pairs is independent of composition, and instead depends solely on the mechanical properties of the parent bodies. This is consistent with the findings of \citet{Pravec10}. 

Taken as a whole our results are most consistent with pair formation via rotational fissioning and/or binary disruption. It is expected that a collisional formation between compositionally distinct bodies would produce at least some primaries and secondaries with disparate colors, though this presumption should be numerically investigated in detail. It is unclear how a collisional formation scenario would influence the color distribution of pairs in Figure \ref{fig.histcomp}. Density differences between C- and S-complex asteroids \citep{Britt02} might be a reason for expecting different pair formation efficiencies from disruptive collisions.

Several avenues for future work would help to further constrain the origin of these objects. Spectra or photometric colors of the components in binary systems could determine whether binary disruption can produce a population of pairs whose primaries and secondaries have similar reflectance properties. New models that address the compositional implications of pair formation via rotational fission and via collisions would be useful. Additional spectroscopic observations (particularly at near-infrared wavelengths) could provide further insight into the composition, extent of weathering and surface properties of these interesting systems.

%% Using an acknowledgements command is not in the Elsevier template,
%% but it can be used.
\ack

I would like to thank Scott Sheppard and Mark Willman for their assistance with observing several of the objects presented here and in their helpful comments on early drafts of this manuscript. Thoughtful comments on the manuscript were also provided by David Polishook. Insightful reviews were kindly provided by David Vokrouhlick\'y and an anonymous referee. This work includes data obtained at the Magellan 6.5m and DuPont 2.5m telescopes located at Las Campanas Observatory in Chile, and at the University of Hawaii 2.2m telescope located on Mauna Kea in Hawaii. Support for this project was provided by the Carnegie Institution of Washington and by the National Aeronautics and Space Administration through the NASA Astrobiology Institute (NAI) under Cooperative Agreement No. NNA04CC09A.

\label{lastpage}

% Bibliographic references with the natbib package:
% Parenthetical: \citep{Bai92} produces (Bailyn 1992).
% Textual: \citet{Bai95} produces Bailyn et al. (1995).
% An affix and part of a reference:
%   \citep[e.g.][Ch. 2]{Bar76}
%   produces (e.g. Barnes et al. 1976, Ch. 2).-

%\bibliography{bibliography.bib}

%% Use the plainnat style for ``Icarus'' mode to display DOI numbers
%% among other things.  However, revert to the Elsevier elsart-harv
%% mode for ``Elsevier'' mode.
%\bibliographystyle{plainnat}
% \bibliographystyle{elsart-harv}

%% --Tables-- 

\clearpage	% Make sure things don't run together.

% OBSERVING CIRCUMSTANCES

\begin{center}
\begin{singlespacing}
{\scriptsize
\begin{longtable}{llccc}
\hline 
\hline
	 					&			& $\Delta$ 	& $R$		& $\alpha$ \\
Object 					& UT Date 	& (AU) 		& (AU)		& (deg) \\
\hline
(2110) Moore-Sitterly 		& 2010-02-28	& 1.661		& 2.513		& 14.2 \\
(15107) Toepperwein		& 2010-02-28	& 1.766		& 2.587		& 14.9 \\
(17288) 2000 NZ10			& 2010-03-08	& 2.071		& 2.488		& 22.9 \\
(17288) 2000 NZ10			& 2010-08-31	& 1.992		& 2.144		& 27.9 \\
(21930) 1999 VP61			& 2010-02-27	& 3.670		& 4.636		& 3.1 \\
(22647) Levi-Strauss		& 2010-03-07	& 3.775		& 4.761		& 1.5 \\
(32957) 1996 HX20			& 2010-08-31	& 2.044		& 2.330		& 25.6 \\
(38395) 1999 RR193		& 2010-08-29	& 3.443		& 3.722		& 15.6 \\
(51609) 2001 HZ32			& 2011-03-05 	& 1.669		& 2.654		& 3.2 \\
(54041) 2000 GQ113		& 2010-02-28 	& 1.774		& 2.611		& 14.1 \\
(69142) 2003 FL115			& 2010-08-30	& 1.903		& 2.041		& 29.4 \\
(70511) 1999 TL103			& 2010-08-29	& 1.404		& 2.373		& 8.9 \\
(84203) 2002 RD133		& 2010-08-29	& 1.069		& 1.850		& 26.3 \\
(92652) 2000 QX36			& 2010-03-08	& 2.000		& 2.465		& 22.8 \\
(99052) 2001 ET15			& 2010-03-08	& 1.666 		& 2.548		& 12.7 \\
(117306) 2004 VF21		& 2010-03-08	& 1.952		& 2.509		& 21.4 \\
(127502) 2002 TP59		& 2012-02-14	& 0.928		& 1.888		& 10.1 \\
(139537) 2001 QE25		& 2010-08-31	& 1.751		& 2.527		& 17.7 \\
(141513) 2002 EZ93		& 2010-08-30	& 3.656		& 3.732		& 15.7 \\
(143662) 2003 SP84		& 2010-02-26 	& 1.081		& 1.950		& 18.8 \\
(143662) 2003 SP84		& 2010-02-27 	& 1.085		& 1.951		& 19.0 \\
(189994) 2004 GH33		& 2010-03-07	& 2.059		& 2.390		& 24.3 \\
(194083) 2001 SP159		& 2010-03-06	& 1.322		& 2.280		& 8.4 \\
(195479) 2002 GX130		& 2010-03-08	& 1.550		& 2.516		& 6.7 \\
(195479) 2002 GX130		& 2010-02-28	& 1.587		& 2.512		& 10.2 \\
(203489) 2002 AL80		& 2010-03-07	& 1.905		& 2.608		& 18.1 \\
(216177) 2006 TE23		& 2010-08-22 	& 1.473		& 2.393		& 12.9 \\
(220143) 2002 TO134		& 2011-03-04 	& 1.747		& 2.624		& 12.4 \\
(229991) 2000 AH207		& 2010-03-01	& 1.133		& 2.088		& 9.8 \\
(237517) 2000 SP31		& 2010-03-06	& 1.798		& 2.709		& 10.3 \\
(279865) 2001 HU24		& 2010-03-06	& 1.755		& 2.529		& 16.9 \\
(284765) 2008 WK70		& 2010-03-06	& 1.739		& 2.547		& 15.7 \\
(291188) 2006 AL54		& 2010-03-01	& 1.396		& 2.256		& 16.0 \\
(291788) 2006 KM53		& 2010-03-07	& 1.628		& 2.581		& 7.6 \\
(303284) 2004 RJ294		& 2010-03-08	& 1.334		& 2.086		& 22.4 \\
2008 TS51				& 2010-03-07	& 1.797		& 2.699		& 10.8 \\
\hline
\hline
\caption[]{\normalsize Asteroid Pairs Observed in this Study\\
The columns in this table are: object number and designation, UT date of observation, heliocentric distance ($\Delta$), geocentric distance ($R$), and phase angle ($\alpha$). Data retrieved from the Minor Planet Center website.
}
\label{tab.obs}
\end{longtable}
} %end small font
\end{singlespacing}
\end{center}

%% BVRI Photometry

\begin{center}
\begin{singlespacing}
\begin{landscape}
{\scriptsize
\begin{longtable}{lcllccccccc}
\hline 
\hline
Object 					& 1/2 & Companion 				& UT Date		& Instrument 	& {\it V} 		& $H_V$		& {\it B - V}		& {\it V - R} 		& {\it R - I} 				& $a^*$ \\
\hline
(2110) Moore-Sitterly$^{bc}$ 	& 1	& (44612) 1999 RP27		& 2010-02-28	& SITe2k	& 17.51 $\pm$ 0.09	& 13.6		& 0.89 $\pm$ 0.02	& 0.45 $\pm$ 0.02	& 0.41 $\pm$ 0.02 		& 0.12 $\pm$ 0.02 \\
{\bf (15107) Toepperwein$^{ce}$}& 1 & (291188) 2006 AL54$^b$	& 2010-02-28	& SITe2k	& 18.81 $\pm$ 0.03 	& 14.7		& 0.90 $\pm$ 0.07	& 0.45 $\pm$ 0.03	& 0.32 $\pm$ 0.06 		& 0.09 $\pm$ 0.07 \\
{\bf (17288) 2000 NZ10$^b$}	& 1	& (203489) 2002 AL80$^b$	& 2010-03-08	& IMACS	& 19.03 $\pm$ 0.01	& 14.4		& 0.90 $\pm$ 0.01	& 0.52 $\pm$ 0.01	& 0.37 $\pm$ 0.01 		& 0.11 $\pm$ 0.01 \\
{\bf (17288) 2000 NZ10$^b$}	& 1	& (203489) 2002 AL80$^b$	& 2010-08-31	& SITe2k	& 18.86 $\pm$ 0.02	& 14.5		& 0.91 $\pm$ 0.04	& 0.52 $\pm$ 0.04	& 0.35 $\pm$ 0.05 		& 0.11 $\pm$ 0.04 \\
{\bf (21930) 1999 VP61$^{ce}$} & 1	& (22647) Levi-Strauss		& 2010-02-27	& SITe2k	& 19.46 $\pm$ 0.02	& 13.0		& 0.66 $\pm$ 0.05	& 0.41 $\pm$ 0.04	& 0.36 $\pm$ 0.05 		& -0.11$\pm$ 0.05 \\
{\bf (22647) Levi-Strauss}		& 2	& (21930) 1999 VP61		& 2010-03-07	& IMACS	& 19.86 $\pm$ 0.04	& 14.2		& 0.76 $\pm$ 0.06	& 0.42 $\pm$ 0.07	& 0.35 $\pm$ 0.09 		& -0.02 $\pm$ 0.07 \\
(32957) 1996 HX20$^c$		& 2	& (38707) 2000 QK89		& 2010-08-31	& SITe2k	& 20.66 $\pm$ 0.04	& 16.1		& 0.86 $\pm$ 0.05	& 0.52 $\pm$ 0.03	& 0.21 $\pm$ 0.06 		& 0.02 $\pm$ 0.05 \\
{\bf (38395) 1999 RR193}		& 1	& (141513) 2002 EZ93		& 2010-08-29	& SITe2k	& 20.13 $\pm$ 0.04	& 13.7		& 0.71 $\pm$ 0.05	& 0.39 $\pm$ 0.07	& 0.39 $\pm$ 0.10 		& -0.05 $\pm$ 0.06 \\
{\bf (51609) 2001 HZ32$^{ce}$} & 1 	& 1999 TE221$^{b}$ 		& 2011-03-05 	& IMACS	& 19.12 $\pm$ 0.06	& 15.6		& 0.82 $\pm$ 0.04	& 0.38 $\pm$ 0.04	& 0.35 $\pm$ 0.07 		& 0.03 $\pm$ 0.05 \\
{\bf (54041) 2000 GQ113$^{bce}$} & 1 & (220143) 2002 TO134 	& 2010-02-28 	& SITe2k	& 19.04 $\pm$ 0.02	& 14.9		& 0.85 $\pm$ 0.08	& 0.50 $\pm$ 0.06	& 0.23 $\pm$ 0.02 		& 0.01 $\pm$ 0.07 \\
{\bf (69142) 2003 FL115$^c$}	& 1	& (127502) 2002 TP59		& 2010-08-30	& SITe2k	& 20.08 $\pm$ 0.19	& 15.9		& 0.88 $\pm$ 0.13	& 0.44 $\pm$ 0.02	& 0.49 $\pm$ 0.07 		& 0.15 $\pm$ 0.12 \\
(70511) 1999 TL103	$^{ce}$	& 1	& 2007 TC334				& 2010-08-29	& SITe2k	& 18.71 $\pm$ 0.04	& 15.5		& 0.93 $\pm$ 0.05	& 0.54 $\pm$ 0.03	& 0.09 $\pm$ 0.05 		& 0.03 $\pm$ 0.05 \\
(84203) 2002 RD133$^c$	& 1	& (285637) 2000 SS4		& 2010-08-29	& SITe2k	& 19.27 $\pm$ 0.07	& 16.6		& 0.67 $\pm$ 0.04	& 0.41 $\pm$ 0.04	& 0.41 $\pm$ 0.05 		& -0.08 $\pm$ 0.04 \\
{\bf (92652) 2000 QX36}		& 1	& (194083) 2001 SP159		& 2010-03-08	& IMACS	& 20.42 $\pm$ 0.01	& 15.9		& 0.88 $\pm$ 0.02	& 0.50 $\pm$ 0.01	& 0.39 $\pm$ 0.01 		& 0.10 $\pm$ 0.02 \\
{\bf (99052) 2001 ET15$^{ae}$} & 1	& (291788) 2006 KM53		& 2010-03-08	& IMACS	& 19.47 $\pm$ 0.01	& 15.6		& 0.87 $\pm$ 0.01	& 0.48 $\pm$ 0.01	& 0.45 $\pm$ 0.01 		& 0.12 $\pm$ 0.02 \\
{\bf (117306) 2004 VF21}		& 2	& (10123) Figeoja$^a$		& 2010-03-08	& IMACS	& 20.91 $\pm$ 0.01	& 16.4		& 0.87 $\pm$ 0.03	& 0.53 $\pm$ 0.01	& - 					& - \\
(139537) 2001 QE25$^c$	& 1	& (210904) 2001 SR218		& 2010-08-31	& SITe2k	& 19.51 $\pm$ 0.04	& 15.4		& 0.67 $\pm$ 0.03	& 0.43 $\pm$ 0.03	& 0.32 $\pm$ 0.06 		& -0.11 $\pm$ 0.04 \\
{\bf (141513) 2002 EZ93$^c$}	& 2	& (38395) 1999 RR193		& 2010-08-30	& SITe2k	& 21.12 $\pm$ 0.10	& 14.6		& 0.68 $\pm$ 0.04	& 0.36 $\pm$ 0.05	& 0.40 $\pm$ 0.08 		& -0.07 $\pm$ 0.05 \\
(143662) 2003 SP84$^c$		& 2 	& (92336) 2000 GY81 		& 2010-02-26 	& SITe2k	& 19.51 $\pm$ 0.13	& 16.9		& 0.88 $\pm$ 0.07	& 0.48 $\pm$ 0.03	& 0.39 $\pm$ 0.04 		& 0.10 $\pm$ 0.07 \\
(143662) 2003 SP84		& 2 	& (92336) 2000 GY81 		& 2010-02-27 	& SITe2k	& 19.45 $\pm$ 0.04	& 16.9		& 0.82 $\pm$ 0.07	& 0.42 $\pm$ 0.04	& 0.47 $\pm$ 0.06 		& 0.09 $\pm$ 0.06 \\
{\bf (189994) 2004 GH33}		& 1	& (303284) 2004 RJ294		& 2010-03-07	& IMACS	& 21.93 $\pm$ 0.05	& 17.3		& 0.89 $\pm$ 0.09	& 0.49 $\pm$ 0.08	& 0.36 $\pm$ 0.10 		& 0.10 $\pm$ 0.09 \\
{\bf (194083) 2001 SP159}	& 2	& (92652) 2000 QX36		& 2010-03-06	& IMACS	& 20.05 $\pm$ 0.05	& 17.1		& 0.84 $\pm$ 0.07	& 0.50 $\pm$ 0.07	& 0.43 $\pm$ 0.08 		& 0.08 $\pm$ 0.07 \\
{\bf (195479) 2002 GX130$^e$} & 1	& (284765) 2008 WK70		& 2010-03-08	& IMACS	& 19.90 $\pm$ 0.01	& 16.4		& 0.88 $\pm$ 0.02	& 0.51 $\pm$ 0.01	& 0.30 $\pm$ 0.01 		& 0.06 $\pm$ 0.02 \\
{\bf (195479) 2002 GX130$^c$}& 1 & (284765) 2008 WK70		& 2010-02-28	& SITe2k	& 20.12 $\pm$ 0.07	& 16.5		& 0.87 $\pm$ 0.08	& 0.43 $\pm$ 0.03	& 0.35 $\pm$ 0.03 		& 0.07 $\pm$ 0.08 \\
{\bf (203489) 2002 AL80$^b$}	& 2	& (17288) 2000 NZ10$^b$	& 2010-03-07	& IMACS	& 21.00 $\pm$ 0.04	& 16.6		& 0.93 $\pm$ 0.07	& 0.50 $\pm$ 0.07	& 0.36 $\pm$ 0.09 		& 0.13 $\pm$ 0.08 \\
{\bf (216177) 2006 TE23$^{cde}$}& 2 & (34380) 2000 RV55$^{b}$ 	& 2010-08-22 	& IMACS	& 20.20 $\pm$ 0.07	& 16.7		& 0.81 $\pm$ 0.05	& 0.46 $\pm$ 0.02	& 0.33 $\pm$ 0.04 		& 0.02 $\pm$ 0.04 \\
{\bf (220143) 2002 TO134$^{ce}$}& 2 & (54041) 2000 GQ113$^{b}$ 	& 2011-03-04 	& IMACS	& 20.87 $\pm$ 0.07	& 16.8		& 0.86 $\pm$ 0.07	& 0.52 $\pm$ 0.05	& 0.28 $\pm$ 0.08		& 0.04 $\pm$ 0.07 \\
(229991) 2000 AH207$^c$	& 2	& (106598) 2000 WZ112		& 2010-03-01	& SITe2k	& 19.65 $\pm$ 0.04	& 17.1		& 0.86 $\pm$ 0.13	& 0.41 $\pm$ 0.06	& 0.30 $\pm$ 0.05 		& 0.05 $\pm$ 0.12 \\
(237517) 2000 SP31		& 1	& 2007 TN127				& 2010-03-06	& IMACS	& 22.22 $\pm$ 0.06	& 18.1		& 0.84 $\pm$ 0.12	& 0.32 $\pm$ 0.08	& 0.25 $\pm$ 0.09 		& 0.01 $\pm$ 0.11 \\
(279865) 2001 HU24		& 2	& (52773) 1998 QU12		& 2010-03-06	& IMACS	& 21.78 $\pm$ 0.05	& 17.6		& 0.88 $\pm$ 0.08	& 0.46 $\pm$ 0.07	& 0.44 $\pm$ 0.09 		& 0.12 $\pm$ 0.08 \\
{\bf (284765) 2008 WK70}	& 2	& (195479) 2002 GX130		& 2010-03-06	& IMACS	& 21.34 $\pm$ 0.05	& 17.3		& 0.78 $\pm$ 0.09	& 0.45 $\pm$ 0.07	& 0.42 $\pm$ 0.09 		& 0.02 $\pm$ 0.09 \\
{\bf (291188) 2006 AL54$^{bc}$}& 2	& (15107) Toepperwein		& 2010-03-01	& SITe2k	& 20.70 $\pm$ 0.08	& 17.3		& 0.86 $\pm$ 0.15	& 0.51 $\pm$ 0.07	& 0.41 $\pm$ 0.06 		& 0.10 $\pm$ 0.14 \\
{\bf (291788) 2006 KM53}		& 2	& (99052) 2001 ET15$^a$	& 2010-03-07	& IMACS	& 20.23 $\pm$ 0.04	& 16.6		& 0.82 $\pm$ 0.06	& 0.48 $\pm$ 0.07	& 0.37 $\pm$ 0.09 		& 0.04 $\pm$ 0.07 \\
{\bf (303284) 2004 RJ294$^b$}& 2	& (189994) 2004 GH33 		& 2010-03-08	& IMACS	& 21.33 $\pm$ 0.02	& 18.0		& 0.90 $\pm$ 0.04	& 0.54 $\pm$ 0.02	& 0.40 $\pm$ 0.01 		& 0.12 $\pm$ 0.04 \\
2008 TS51				& 2	& 2002 RJ126				& 2010-03-07	& IMACS	& 22.49 $\pm$ 0.06	& 18.4		& 0.66 $\pm$ 0.11	& 0.29 $\pm$ 0.08	& 0.28 $\pm$ 0.11 		& -0.14 $\pm$ 0.11 \\
\hline
\hline
\caption[]{\normalsize Summary of Asteroid Pair {\it BVRI} observations\\
The columns in this table are: object number and designation, pair primary (1) or secondary (2), companion number and designation, UT date of observation, instrument employed, measured {\it V}-band magnitude, calculated absolute $V$-band magnitude at zero phase angle (see text), and photometric colors. Entries in bold indicate pairs where data was obtained or already exists for both components.\\
$^a$ Observed spectroscopically.\\
$^b$ Observed by SDSS.\\
$^c$ Multiple {\it BVRI} sequences. The V-band error represents the standard deviation of all measured V-band magnitudes. \\
$^d$ Spurious pair.\\
$^e$ Photometry calibrated with SDSS field stars.
}
\label{tab.bvri}
\end{longtable}
} %end small font
\end{landscape}
\end{singlespacing}
\end{center}

%% TABLE OF OBJECTS WITH ugriz PHOTOMETRY
%% griz Photometry
\begin{center}
\begin{singlespacing}
\begin{landscape}
{\scriptsize
\begin{longtable}{lclccccccc}
\hline 
\hline
Object 					& 1/2 & Companion 				& {\it r} 			& {\it u - g} 		& {\it g - r} 			& {\it r - i} 			& {\it i - z}				& $a^*$ \\
\hline
(2110) Moore-Sitterly$^a$ 	& 1	& (44612) 1999 RP27		& 14.88 $\pm$ 0.03	& 1.73 $\pm$ 0.06	& 0.67 $\pm$ 0.04	& 0.28 $\pm$ 0.04	& -0.05 $\pm$ 0.03 		& 0.15 $\pm$ 0.04 \\
(4765) Wasserburg			& 1	& 2001 XO105				& 16.46 $\pm$ 0.02	& 1.41 $\pm$ 0.04	& 0.49 $\pm$ 0.04	& 0.15 $\pm$ 0.02	& 0.05 $\pm$ 0.02 		& -0.07 $\pm$ 0.04 \\
(10484) Hecht				& 1	& (44645) 1999 RC118		& 17.07 $\pm$ 0.01	& 1.76 $\pm$ 0.03	& 0.39 $\pm$ 0.01	& 0.20 $\pm$ 0.01	& -0.26 $\pm$ 0.02 		& 0.16 $\pm$ 0.01 \\
{\bf (17288) 2000 NZ10$^a$}	& 1	& (203489) 2002 AL80$^a$	& 17.87 $\pm$ 0.01	& 1.66 $\pm$ 0.16	& 0.68 $\pm$ 0.02	& 0.17 $\pm$ 0.02	& -0.01 $\pm$ 0.03 		& 0.11 $\pm$ 0.02 \\
{\bf (34380) 2000 RV55$^c$}	& 1	& (216177) 2006 TE23$^a$	& 19.60 $\pm$ 0.02	& 1.71 $\pm$ 0.27	& 0.72 $\pm$ 0.04	& 0.24 $\pm$ 0.04	& -0.07 $\pm$ 0.06 		& 0.18 $\pm$ 0.04 \\
(38184) 1999 KF			& 1	& (221867) 2008 GR90		& 18.77 $\pm$ 0.02	& 1.70 $\pm$ 0.12	& 0.70 $\pm$ 0.03	& 0.18 $\pm$ 0.03	& -0.01 $\pm$ 0.04 		& 0.13 $\pm$ 0.03 \\
{\bf (54041) 2000 GQ113$^a$} & 1 	& (220143) 2002 TO134$^a$	& 18.27 $\pm$ 0.02	& 1.58 $\pm$ 0.06	& 0.63 $\pm$ 0.04	& 0.20 $\pm$ 0.03	& -0.20 $\pm$ 0.03 		& 0.08 $\pm$ 0.04 \\
(63440) 2001 MD30			& 1	& 2004 TV14				& 17.10 $\pm$ 0.01	& 1.29 $\pm$ 0.07	& 0.52 $\pm$ 0.02	& 0.17 $\pm$ 0.01	& 0.02 $\pm$ 0.15 		& -0.03 $\pm$ 0.02 \\
(106700) 2000 WX167 		& 1	& (263114) 2007 UV			& 19.07 $\pm$ 0.02	& 1.36 $\pm$ 0.08	& 0.59 $\pm$ 0.03	& 0.15 $\pm$ 0.03	& 0.08 $\pm$ 0.04 		& 0.02 $\pm$ 0.03 \\
(113029) Priscus			& 1	& (113029) 2002 RZ46		& 19.93 $\pm$ 0.02	& 1.07 $\pm$ 0.30	& 0.69 $\pm$ 0.04	& 0.26 $\pm$ 0.04	& -0.05 $\pm$ 0.08 		& 0.16 $\pm$ 0.04 \\
{\bf (127502) 2002 TP59$^b$}	& 2	& (69142) 2003 FL115$^a$	& 18.76 $\pm$ 0.10	& -				& 0.57 $\pm$ 0.15	& 0.22 $\pm$ 0.11	&  - 			 		& 0.04 $\pm$ 0.14 \\
{\bf (203489) 2002 AL80$^a$}	& 2	& (17288) 2000 NZ10$^a$	& 20.02 $\pm$ 0.02	& 1.06 $\pm$ 0.30	& 0.65 $\pm$ 0.04	& 0.19 $\pm$ 0.04	& -0.04 $\pm$ 0.08 		& 0.09 $\pm$ 0.04 \\
(250322) 2003 SC7			& 2	& (52852) 1998 RB75		& 19.91 $\pm$ 0.02	& 0.87 $\pm$ 0.35	& 0.58 $\pm$ 0.04	& 0.19 $\pm$ 0.04	& -0.10 $\pm$ 0.09 		& 0.03 $\pm$ 0.04 \\
{\bf (291188) 2006 AL54$^a$}	& 2	& (15107) Toepperwein$^a$	& 19.88 $\pm$ 0.02	& 1.39 $\pm$ 0.20	& 0.67 $\pm$ 0.04	& 0.20 $\pm$ 0.03	& 0.00 $\pm$ 0.06 		& 0.11 $\pm$ 0.04 \\
{\bf (303284) 2004 RJ294$^a$}& 2	& (189994) 2004 GH33$^a$	& 21.09 $\pm$ 0.05	& 1.45 $\pm$ 0.54 	& 0.69 $\pm$ 0.08	& 0.16 $\pm$ 0.07	& -0.10 $\pm$ 0.20 		& 0.12 $\pm$ 0.08 \\
{\bf 1999 TE221}			& 2	& (51609) 2001 HZ32$^a$	& 18.66 $\pm$ 0.02	& 1.46 $\pm$ 0.08	& 0.55 $\pm$ 0.03	& 0.20 $\pm$ 0.03	& -0.32 $\pm$ 0.04 		& 0.01 $\pm$ 0.03 \\
2003 YK39				& 2	& (21436) Chaoyichi			& 19.82 $\pm$ 0.03	& 1.92 $\pm$ 0.31	& 0.75 $\pm$ 0.04	& 0.06 $\pm$ 0.04	& -0.07 $\pm$ 0.07 		& 0.12 $\pm$ 0.04 \\
\hline
\hline
\caption[]{\normalsize Summary of Asteroid Pair {\it ugriz} data\\
In all but one case, (127502) 2002 TP59, these data were retreived from the SDSS MOC. The columns in this table are: object number and designation, pair primary (1) or secondary (2), companion number and designation, $r$-band magnitude, and photometric colors. Entries in bold indicate pairs where data was obtained for both components.\\
$^a$ BVRI observations were performed.\\
$^b$ Observed with SNIFS at the University of Hawaii 2.2m telescope on UT 14 February, 2012. Multiple {\it gri} sequences were obtained. The phase corrected absolute magnitude for this object is $H_r$=16.9 (see text). \\
$^c$ Spurious pair.
}
\label{tab.ugriz}
\end{longtable}
} %end small font
\end{landscape}
\end{singlespacing}
\end{center}

%% TABLE OF SPECTROSCOPIC OBSERVATIONS SUMMARY
\begin{center}
\begin{table}[]
\begin{tabular}{llcclc}
\hline 
\hline
Object 			& UT Date 	& Mag. 	& $t_{exp}$ (s) 	& Solar Analog		& Taxon \\
\hline
(10123) Figeoja	& 2010-03-06	& 17.7	& 600		& SA104-335		& Ld \\
(99052) 2001 ET15	& 2010-03-08	& 19.47	& 600		& HD127913		& S \\
\hline
\hline
\end{tabular}
\caption{IMACS spectroscopic observations}
\label{tab.spec}
\end{table}%
\end{center}

%% TABLE OF a* VALUES AND TAXA
\begin{center}
\begin{table}[]
\begin{tabular}{llccc}
\hline 
\hline
	 				& 		 			& Primary 		 	& Secondary 		& Taxonomic  \\
Primary 				& Secondary 			& $a^*$ 			& $a^*$			& Complex \\
\hline
(15107) Toepperwein	& (291188) 2006 AL54	& 0.09 $\pm$ 0.07 	& 0.09 $\pm$ 0.14	& S \\
(17288) 2000 NZ10		& (203489) 2002 AL80	& 0.11 $\pm$ 0.01 	& 0.13$\pm$ 0.08	& S  \\
(21930) 1999 VP61		& (22647) Levi-Strauss	& -0.11$\pm$ 0.05 	& -0.02$\pm$ 0.07	& C \\
(38395) 1999 RR193	& (141513) 2002 EZ93	& -0.05 $\pm$ 0.06 	& -0.07$\pm$ 0.05	& C \\
(51609) 2001 HZ32		& 1999 TE221		 	& 0.03 $\pm$ 0.05 	& 0.01 $\pm$ 0.03	& S \\
(54041) 2000 GQ113	& (220143) 2002 TO134 	& 0.01 $\pm$ 0.07 	& 0.04$\pm$ 0.07	& S \\
(69142) 2003 FL115		& (127502) 2002 TP59	& 0.15 $\pm$ 0.12	& 0.04$\pm$ 0.14	& S  \\
(92652) 2000 QX36		& (194083) 2001 SP159	& 0.10 $\pm$ 0.02	& 0.08$\pm$ 0.07	& S  \\
(99052) 2001 ET15		& (291788) 2006 KM53	& 0.12 $\pm$ 0.01 	& 0.04$\pm$ 0.07	& S \\
(189994) 2004 GH33	& (303284) 2004 RJ294	& 0.10 $\pm$ 0.09	& 0.12$\pm$ 0.04	& S  \\
(195479) 2002 GX130	& (284765) 2008 WK70	& 0.06 $\pm$ 0.02 	& 0.02 $\pm$ 0.05	& S \\
\hline
\hline
\end{tabular}
\caption{Comparison of Primary and Seconday $a^*$ Colors}
\label{lasttable}
\label{tab.comp}
\end{table}%
\end{center}

\clearpage

%% --Figures-- %%

%% Pair 99052-291788

%% PCA Analysis

%% Pair 69142-127502

%% Pair 10123-117306

%% Pair 34380-216177

\end{document}